\documentclass[11pt]{article}
\usepackage[left=2cm, right=2cm, top=2.5cm, bottom=2.5cm]{geometry}
\usepackage{float}

\usepackage[x11names,dvipsnames]{xcolor}
\usepackage{fancyhdr, amssymb, amsmath, graphicx, pgfplots, tikz}
\usepackage{isomath}

\usetikzlibrary{shadows}

\newcommand{\scolor}{Blue}

\usepackage[labelfont={bf,sf, color=\scolor}, margin={1.5cm,0cm}]{caption}
\usepackage[colorlinks=true, urlcolor=\scolor!80!white, linkcolor=\scolor, citecolor=\scolor!80!white, hyperindex=true, linktocpage=true]{hyperref}

\usepackage[explicit]{titlesec}
\def\L{\mathcal{L}}
\def\p{\partial}  
\def\cp{\mathbb{CP}^2}
\def\a{\alpha}
\def\lag{\langle}
\def\rag{\rangle}
\def\hri#1#2{\href{http://arxiv.org/abs/#1}{[ArXiv:#1]#2}}

\titleformat{\section}
{\gdef\sectionlabel{}
	\Large\bfseries\scshape}
{\gdef\sectionlabel{\thesection }}{0pt}
{\begin{tikzpicture}[remember picture]
	\draw (0.2, 0) node[right] {\color{\scolor} \textsf{#1}};
	\draw (0.0, 0) node[left, fill=MidnightBlue,circle] {\color{white} \textsf{\sectionlabel}};
	\end{tikzpicture}
}
\titlespacing*{\section}{0pt}{20pt}{5pt}
	\newcommand*\subsectionlabel{}
	\titleformat{\subsection}
	{\gdef\subsectionlabel{}
		\large\bfseries\scshape}
	{\gdef\subsectionlabel{\thesubsection  }}{0pt}
	{\begin{tikzpicture}[remember picture]
		\draw (0.15, 0) node[right] {\color{\scolor}\textsf{#1}}; 
		\draw[blue] (0.15, 0) node[left,fill=CornflowerBlue , circle]{ {\color{white} \textsf{\subsectionlabel}}};
		\end{tikzpicture}
	}
	\titlespacing*{\subsection}{5pt}{10pt}{0pt}

\begin{document} 
	\allowdisplaybreaks
	
	\pagestyle{fancy}
	\renewcommand{\headrulewidth}{0pt}
	\fancyhead{}
	
	\fancyfoot{}
	\fancyfoot[C] {\textsf{\textbf{\thepage}}}
\begin{equation*}
	\begin{tikzpicture}
	\draw (0.5\textwidth, -3) node[text width = \textwidth] { \huge \textsf{\textbf{Non-Equilibrium Critical Phenomena From Probe Brane Holography in Schr\"odinger Spacetime}} };
	\end{tikzpicture}
\end{equation*}

\begin{equation*}
	\begin{tikzpicture}
	\draw (0.5\textwidth, 0.1) node[text width=\textwidth] {\large \color{black}  \textsf{{\color{\scolor}Ali Vahedi} and {\color{\scolor} Mobin Shakeri}}};
	\draw (0.5\textwidth, -0.5) node[text width=\textwidth] {\small\textsf{Departemet of Physics, Kharazmi University, Mofatteh Ave, Tehran, Iran}};
	\draw (0.5\textwidth, -1.0) node[text width=\textwidth] {\small\textsf{Applied Science Research Center(ASRC), Kharazmi University, Karaj, Iran}};
	\end{tikzpicture}
\end{equation*}
\begin{equation*}
	\begin{tikzpicture}
	\draw (0, -13.1) node[right, text width=0.5\paperwidth] {\texttt{vahedi@khu.ac.ir , vahedi@ipm.ir mobin.shakeri@live.com}};
	\draw (\textwidth, -13.1) node[left] {\textsf{\today}};
	\end{tikzpicture}
\end{equation*}
\begin{equation*}
	\begin{tikzpicture}
	\draw[very thick, color=\scolor] (0.0\textwidth, -5.75) -- (0.99\textwidth, -5.75);
	\draw (0.12\textwidth, -6.25) node[left] {\color{\scolor}  \textsf{\textbf{Abstract:}}};
	\draw (0.53\textwidth, -6) node[below, text width=0.8\textwidth, text justified] {\small  We study the non-equilibrium steady-state phase transition from probe brane holography in $z=2$ Schr\"odinger spacetime. Concerning differential conductivity, a phase transition could occur in the conductor state.  Considering constant current operator as the external field and the conductivity as an order parameter, we derive scaling behavior of order parameter near the critical point. We explore the critical exponents of the non-equilibrium phase transition in two different Schr\"odinger spacetimes, which originated $1)$ from supergravity, and $2)$ from AdS blackhole in the light-cone coordinates. 
		 Interestingly, we will see that even at the zero charge density, in our first geometry, the dynamical critical exponent of $z=2$ has a major effect on the critical exponents. };
	\end{tikzpicture}
\end{equation*}

\tableofcontents

\begin{equation*}
	\begin{tikzpicture}
	\draw[very thick, color=\scolor] (0.0\textwidth, -5.75) -- (0.99\textwidth, -5.75);
	\end{tikzpicture}
\end{equation*}

\section{Introduction}
\label{sec:intro}
Believing the holographic principle, holographic probe branes give us an analytic description of the non-linear electric conductivity of strongly coupled  QFTs~\cite{kobayashi06,KO07}. The $N_f$ $D7$-branes as a probe, in the background of $N_c$ $D3$-branes introduce the fermions(and anti-fermions) in the dual theory with fundamental(and anti-fundamental) representation of supersymmetric $\mathcal{N}=2$, which localized on a defect sector within the background plasma, $\mathcal{N}=4$ supersymmetric Yang-Mills theory~\cite{Karch02}. The null Melvin twist(NMT) transformation of this configuration, will produce the extension of the holographic probe branes on the Schr\"odinger background~\cite{Am:10}. It was convincing that Schr\"odinger geometry, in general, is dual to the theory of cold fermions at unitarity \cite{Son:08,Balasubramanian:08} \footnote{As mentioned in \cite{Am:10}, Type IIB string theory in Schr\"odinger background is not exactly a dual to the theory of fermions at unitarity. Although, the correlation functions, especially the three-point function, from holographic approach are in agreement with the three-point function of fermions at unitarity, see for example~\cite{Fuertes:2009ex,Volovich:2009yh}.}. As well as the NMT transformation, it was shown that the isometry of AdS geometry in the light-like coordinate will also reduce to the Sch\"odinger group~\cite{Maldacena:2008wh,Kim:10}. Holographic probe branes in the AdS metric at the light-cone frame(ALCF), for the non-zero charge density, might establish a framework to study Strange metal \cite{Kim2:10,kazem1}. Beside the two form B-field, at zero temperature, both Shr\"odinger background from ALCF and NMT transformation have the metric with the $z=2$ dynamical critical exponent. The B-field will break the SUSY in the solution derived from NMT procedure, therefore, in general, they have different dual theories. Note that the QFT with Schr\"odinger symmetry which created by the above transformations lives in one dimension lower than its original QFT with conformal symmetry.
\\
By turning an electric field on the probe D-brane, the world-volume horizon will emerge and we can assign a temperature to this horizon. This temperature, is generally distinct from the background temperature. This means that (anti-)particles in the (anti-)fundamental representation dual theory live in a state with a different temperature from the background plasma(adjoint representation). Therefore, the system is standing at a non-equilibrium steady state, see~\cite{Kundu:13,Kundu1:15,Kundu2:16} for detailed discussion. In other words, the electric field pumps energy into the fermions(flavors) sector and then it dissipates into the background plasma, see also~\cite{Hashimoto:2010wv}.Hence, the system is in out of equilibrium.
Non-linear differential conductivity of systems in a non-equilibrium steady state is studied from holographic probe branes in the $AdS$ background at~\cite{Nakamura:10,Nakamura:12,Ali:13}. It was shown that the sign of the differential conductivity could be changed from negative to positive. The continuous (second order), first-order phase transitions and crossover from negative differential conductivity(NDC) to positive differential conductivity (PDC) are studied in~\cite{Nakamura:10,Nakamura:12,Ali:13,Nakamura:18}. Also in~\cite{Nakamura:12,Nakamura:18,Zeng:18}, the order parameters of non-equilibrium steady state at the critical point are proposed and their scalings are investigated.    
\\
According to the bulk theory for other geometries, the mechanism that makes the system out of equilibrium is similar to the AdS background. In this work, we are going to study the non-equilibrium steady state in Schr\"odinger background with $z=2$ dynamical critical exponent. We investigate and compare the NDC to PDC phase transitions in the both Schr\"odinger spacetime, ALCF and NMT. 
\\
\\
\section{Non-linear DC Conductivity From Probe Branes}
\label{sec:sigma}
Consider a strongly correlated QFT with emerging Schr\"odinger symmetry at its critical point and conserved $U(1)$ current $J^a$. The Schr\"odinger symmetry admits Lifshitz scaling
\begin{equation}\label{eq:sc}
t \to \lambda^z\,t \,,\qquad \vec{x}\to \, \lambda \vec{x} ,
\end{equation}
 where $z$ is known as dynamical critical exponent. For $z=2$, the holographic dual to this theory could be $N_f$ $D7$-branes probing the background with null Melvin twist(NMT) of $AdS_5\times S^5$~\cite{Am:10,Va:17} or $AdS_5 \times S^5$ in light-cone frame(ALCF)\cite{Kim2:10}  with non-zero $A_a$  gauge field on the $D7$-branes. As mentioned, the $N_f$ $D7$-branes will be treated as probes in the background of $N_c$ D3-branes. Thus, the dynamics of the system is given by the Dirac-Born-Infeld(DBI) action for $D7$-branes~\footnote{There is also the Wess-Zumino action term, which for our embedding is zero. We also normalized the action with $D3$-branes world-volume.}:
\begin{equation}\label{DBI0}
S_{D7} = -\mathcal{N}\!\!\int\!dr\, \,e^{-\Phi}\sqrt{-{\det}\left(P\left[g+B\right]_{ab}
+ \left(2\pi\a'\right) F_{ab}\right)} \,,
\end{equation} 
where $\mathcal{N}=2\pi^2 N_f T_{D7}$, and $P[g+B]_{ab}$ stands for pullback or induced metric and induced two-form B-field on the D7 branes. The $F_{ab}$ is a gauge field stress tensor on the probe branes. To extract the equations of motion from Eq.\eqref{DBI0}, we assume the following embeddings for the probe $D7$-branes:
 \begin{table}[H]
	\centering
	\begin{tabular}{|c|c|c|c|c|c|c|c|}
		\hline
		&$x^+$&$x^-$&x ,y & r&$\sigma_\a$ or $S^3$&$\theta$&$\chi$\\
		\hline 
		D3&$\times $& $\times $ &$\times $ & $ $&&&\\
		\hline
		D7&$\times $ & $\times $ & $\times $&$\times $&$\times$&&\\
		\hline
	\end{tabular}
	\caption{\label{tab:1} $D3-D7$ embedding.}
\end{table}
\noindent As it is clear, There is $O(2)$ symmetry along $(\theta,\chi)$. Without loss of generality, we make the assumption that $\theta=\theta(r)$ and $\chi=0$. We assume the following gauge fields on the probe branes
\begin{equation}\label{eq:gauge0}
2\pi \a' A_x(x^+,x^-,r)=E_b x^+ -b^2\,E_b\,x^-+a_x(r)\footnote{This is the same gauge field which is introduced to AdS background,i.e.,$E\, t +a_x(r)$, in the light-like coordinates, which $E_b=\frac{E}{2\,b}$.},
\end{equation}
which $E_b$ is a constant non-relativistic electric field. We consider zero charge density\footnote{ for non-zero finite charge density see~\cite{Am:10,Kim2:10,kazem1}.}, so with this ansatz, we have two equations of motion for $A_a$ and $\theta(r)$ to solve:
\begin{equation}\label{eq:gauge2}
\frac{d}{dr} \frac{\p \L}{\p a_x '}=0 \qquad or\qquad \frac{\p \L}{\p a_x '}=D\,,
\end{equation}\label{eq:theta0}
\begin{equation}
\frac{\p \L}{\p\theta}-\frac{d}{dr} \frac{\p \L}{\p \theta'}=0 \,,
\end{equation}
where the prime stands for the derivative relative to $r$. At the near boundary $r=0$ the solutions of Eq.\eqref{eq:gauge2} would be 
\begin{subequations}
\begin{equation}\label{eq:gauge3}
a_x(r)\approx c+\frac{\lag J^x\rag}{(2 \pi \a')^2 \mathcal{N}}\, r^2+\dots\,,
\end{equation}
where from guage-gravity dictionary~\cite{KO07}, $\lag J^x \rag=D$. The near boundary solution of Eq.\eqref{eq:theta0} is  
 \begin{align}
\theta(r) \approx 2\pi \a' m_q \, r+\frac{\lag \bar{\psi}\psi\rag}{ \mathcal{N}} \,r^3+\dots \,
\end{align}
or
\begin{equation}
	2 \pi \a' m_q=\lim_{r\to 0}\frac{\theta(r)}{r}\,
\end{equation}
where $ m_q$ is a flavor's(quark's) mass.
\end{subequations}
From Eq.\eqref{eq:gauge2}, we could find the on-shell action and also Legendre transform of the action, $\tilde{\L}=\L- A'_x \frac{\p \L}{\p A_x '} $.\\ For zero gauge fields on the probe branes, there are two embeddings. Minkowski embedding (ME) and the black hole embedding(BE). In the bulk theory, for the adequately small ratio of flavor's mass and background Hawking temperature i.e., $\frac{m_q}{T}$, the BE is thermodynamically preferred and for the sufficiently large values of $\frac{m_q}{T}$ the ME is favorable~\cite{CasalderreySolana:11,Albash:07}.
\\ For the non-zero electric field $E_b$, larger than the critical value\footnote{Which is order of effective string tension on the probe branes} $E_b^c$, we able to calculate the electric conductivity $\lag J^x\rag=\sigma\,E_b$, from the reality condition of the DBI action. In other words, for $E_b>E_b^{c}$ there would be a non-zero current $\lag J^x\rag$ hence, we have a conductor state and for  $E_b<E_b^{c}$  we live in an insulator state, since $\lag J^x\rag=0$. The phase transitions may happen from the insulator state to the conductor state.
 The presence of the electric field will introduce the world-volume horizon on the probe branes, which in general, differs from the background event horizon. This makes another class of embedding in addition to the Minkowski and the black hole embeddings. Following~\cite{Ali:13}, we call it Minkowski with the horizon embedding(MHE). The  Minkowski embedding(ME), in the bulk side, corresponds to the states with $\lag J^x\rag=0$, which means that the electric field does not have enough strength to rip off the strings and the bound between the charge carrier pairs would be stable. Differently, the states with $\lag J^x\rag\ne 0$ is demonstrated by the BE or the MHE. According to our numerical calculation, we will see that for a fixed electric field there would be two non-zero currents. Therefore, in the conductor state the phase transitions may also happen. In the following sections, we investigate the other phase transitions in conductor states through the study of non-linear DC conductivity for a QFT with $z=2$ Schr\"odinger symmetry.
\section{Probe Branes In Schr\"odinger Spacetime From NMT }
\label{sec:sNMT}
%
 We consider $N_f$ D7 branes in the below background 
\begin{equation}\label{metric0}
	\begin{split}
		ds^2&=\frac{1}{r^2 K}\left[\big(\frac{f}{r^2}-\frac{1-f}{4 b^2}\big)dx^{+2}+\big(1+f\big)dx^+ dx^- +\big(1-f\big)b^2 dx^{-2}\right]\\
		&+\frac{1}{r^2}\left[d\vec{x}+\frac{dr^2}{f}\right]+ \frac{1}{K} \left(d\chi + \mathcal{A} \right)^2 + ds^2_{\cp}\,,
	\end{split}
\end{equation}

\noindent where
\begin{equation}
	f= 1 - \frac{r^4}{r_H^4}\,, \qquad K= 1 + \frac{b^2 r^2}{r_H^4}\, .
\end{equation}
In addition to this metric, there is a dilatonic scalar field,
\begin{equation}
	\Phi = -\frac{1}{2}Log\, K\,,
\end{equation}
and there is also two-form B-field,  
\begin{equation}
	B= - \frac{1}{2 r^2 K} \left(d\chi + \mathcal{A} \right) \wedge \left( \left( 1 + f \right) \, dx^+ + \left( 1 - f \right) 2 b^2  dx^- \right)\,.
\end{equation}
In this spacetime event horizon located at $r_H$ and the boundary at $r=0$. This geometry is holographic dual to the  thermal quantum state which lives in the temperature equal to the background Hawking temperature. The momentum along the $x^-$ in the dual boundary theory, which is discrete, is number operator generator of Schr\"odinger algebra. In the gravity side this means we might have preformed DLCQ along the $x^-$. So the boundary dual theory also 
has a chemical potential $\mu$\footnote{This quantity should not be confused with chemical potential due to $U(1)$ baryon number of flavor fermions.},see~\cite{Am:10}, with
\begin{equation}\label{T:mu}
	T=\frac{1}{\pi \, b\, r_H}\,\, ,\qquad \mu=\dfrac{-1}{2\, b^2}\,.
\end{equation} 
At the zero temperature (or $r_H \to \infty$) the metric of Eq.\eqref{metric0} changes to
\begin{equation}
	\label{met:scal}
	ds^2  = \frac{1}{r^2} \left(- \frac{1}{r^2 } dx^{ +2} + 2 dx^+ dx^- + d\vec{x}^2+ dr^2 \right)+\left(d\chi + \mathcal{A} \right)^2 + ds^2_{\cp}
\end{equation}
which respects the scale invariant as follows
\begin{equation}\label{scale}
	x^+ \to \lambda^2 x^+ \quad  ,\quad \vec{x} \to \lambda \vec{x}\quad , \quad x^- \to \lambda^0 x^-\,,\quad r \to \lambda r\,.
\end{equation} 
Comparing the above metric with scaling of Eq.\eqref{eq:sc}, in here,
 we deal with the dynamical critical exponent $z=2$.
\\
The Legendre transformation of DBI action Eq.\eqref{DBI0} would be\footnote{See~\cite{Va:17}. The Legendre transformation make current $J$ as a controlling parameter~\cite{Nakamura:10}. } 
\begin{equation}\label{leg2}
	S_{D7} =\int \tilde{\L} \, dr= -\int_{0}^{r_H}\!\!dr g_{rr}^{1/2}
	\sqrt{U(r)\, V(r)}.
\end{equation}
where we have defined:
\begin{align}
	g_{rr}&=\frac{1}{r^2 f(r)}+\theta'^2\label{grr}\\
	U(r)&=\frac{ -4 b^2 r^2 E_b^2\left(r^2-b^2 f(r)\sin^2 \theta\right) + f(r)}{r^2\,f(r)}\, \notag\\
	V(r)&=\frac{\mathcal{N}^2 f(r)\cos^6\theta}{64 r^6}-\lag J^x \rag^2\,\, .
\end{align}
$U$ and $V $, in Eq.\eqref{leg2}, can be positive or negative. The reality condition of the action force us to have a special point, $r_*$, which at this point, $U$ and $V$ change their sign, simultaneously. This means that 
\begin{align} 
	V (r_*) &= 0 \to \lag J^x \rag ^2=\frac{\mathcal{N}^2 f(r_*)\cos^6 \theta(r_*)}{64 r_{*}^6}, \label{eq:con1}\\
	U(r_*)&=0\to E_b^2=\frac{f(r_*)}{4\, b^2\,r^{2}_{*}(r^{2}_{*}-b^2 f(r_*)\sin^2\theta(r_*))}\, .
	\label{eq:con2}
\end{align}
 Following open string metric approach to DBI action we could say that the $r_*$ is a location of world-volume (or apparent) horizon~\cite{CasalderreySolana:11,Albash:07,Hashimoto:2010wv,Kundu:13,KP:11} and we could assign this point an effective temperature, see for detail~\cite{Kundu1:15,Kundu2:16}. Also, from Eq.\eqref{eq:con1} and Eq.\eqref{eq:con2}, one might say that we able to assign a geometric meaning to the electric field, see also~\cite{KP:11}. It is clear from Eq.\eqref{eq:con2}, at the zero electric field $r_*=r_H$ therefore, as already discussed, we
 summarize the embeddings as follows 
\begin{equation}
	Embedding\,\, classes =	\left\{ \begin{array}{l l}    
		r_H>r_*>r_\a &\mbox{for ME}\\
		r_H>r_\a>r_* &\mbox{for MEH}\\
		r_H\geq r_* &\mbox{for BE}
	\end{array}\right.
\end{equation}
where $r_\a$ is a shrinking point of the compact coordinates of the probe branes. 
In the ME the flavor pairs bound is stable, which means current $\lag J^x \rag$ is zero hence the system lives as an insulator. The non-zero current exist for both MEH and BE. Consequently, they are signs of conductor state of a system. In the following, we show that the phase transitions could occur in the conductor state.\\
Before going further, let us focus on Eq.\eqref{eq:con1} and Eq.\eqref{eq:con2}. We could simply drive a formula for the conductivity\footnote{In general, because of the square root, there is a $\pm$ for the conductivity, but without lose of generality, we pick the positive sign.}
\begin{equation}\label{sigma0}
	\lag J^x \rag=\sigma \, E_b \to \sigma= \frac{\mathcal{N}\, b \cos^3\theta(r_*)}{4 r_*^2 }\sqrt{r^{2}_*-b^2 \sin^2\theta(r_*) f(r_*)}\, \, .
\end{equation}
This is non-linear DC conductivity, since $r_*$ and $\theta(r_*)$ are functions of $E_b$(and $\lag J^x \rag$), which can be inferred from Eq.\eqref{eq:con1} and Eq.\eqref{eq:con2}. This is a dimensionless quantity as it was expected to be, in non-relativistic $z=2$  theory~\cite{Am:10}. As mentioned before, the conducting states exist for BE and MEH. To illustrate this statement we focus on $E-J$( V-I or Ohm's law) plot by solving the Euler-Lagrangian equation numerically for $\theta(r)$, in the next section.

\subsection{Realization of Two States: BH and MEH}
\begin{figure}[h!]
	\makebox[\textwidth][c]{\includegraphics[width=1.2\textwidth]{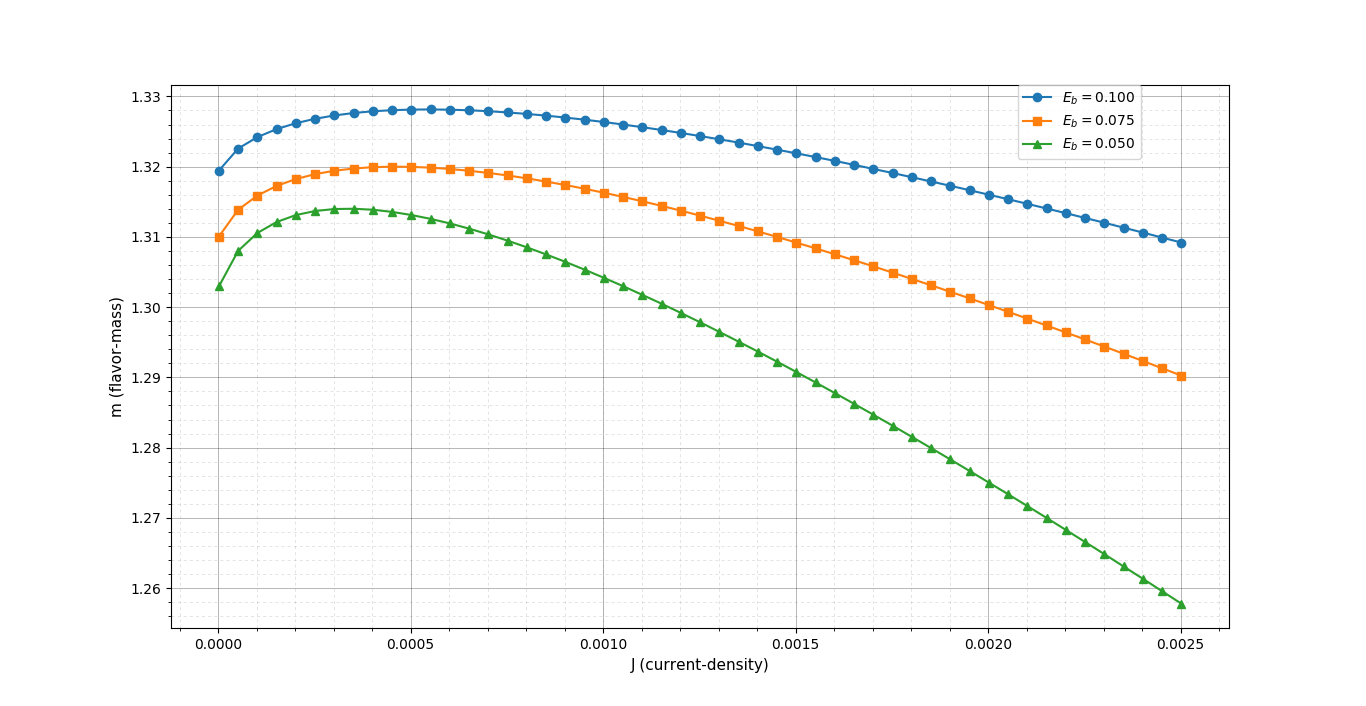}}%
	\caption{$m_q$-$J$ curve for different values of $ E_b$ at $T=0.45015 $ and $b=1$. the maximum values are located at $m_{max}=1.32813$, $1.32000$, and $1.31401$; from top to bottom respectively.}
	\label{fig:1}
\end{figure}

In the AdS background, it was shown that the  Euler-Lagrange equation could be solved by the initial conditions at the point $r_*$~\cite{Nakamura:10,Nakamura:12,Ali:13,Nakamura:18}. To solve the second order equation Eq.\eqref{theta0}, we need two initial or boundary conditions.
Following \cite{Nakamura:10,Nakamura:12}, we choose our initial conditions at $\theta(r_*)$ and $\frac{d \theta}{dr}(r_*)=\theta'(r_*)$. For the $\theta(r_*)$, we pick  one value from $0<\theta(r_*)<\pi/2$. To choose the right value for $\theta'(r_*)$ we expand $\theta(r)$ near the $r_*$ as
\begin{equation}\label{theta0}
\theta(r)=\theta(r_*)+(r-r_*)\theta'(r_*)+\dots\quad .
\end{equation}
Inserting Eq.\eqref{theta0} into the Euler-Lagrange Equation, with the help of Eq.\eqref{eq:con1} and Eq.\eqref{eq:con2}, we will find $\theta'(r_*)$\footnote{Which due to the long and cumbersome expressions which lead to it, we avoid writing it explicitly.}. Now we can solve the Euler-Lagrange Equation, and from the solution we are able to read the mass of flavors from  
\begin{equation}
2\pi \a' m_q=\lim_{r\to 0} \frac{\theta(r)}{r} \, .
\end{equation}
With a fixed external electric field $E_b$, we are able to draw the relation of the flavor's mass to the current $J$. The result is Fig.\eqref{fig:1}, for a fixed background temperature, and a fixed chemical potential\footnote{For the sake of simplicity, from numerical viewpoint and from \eqref{T:mu}, we use $b$  instead of chemical potential.}.
From Fig.\eqref{fig:1}, we find the same behavior as $m-J$ plot in the AdS background \cite{Nakamura:10,Nakamura:12,Ali:13}.

 There is a region where for one value of flavor's mass, we have two different values for current $J$. This is the same statement as saying the both embedding, MEH and BE, could exist for non-zero current. We also, see that for example, at $E=0.05$, for $m>1.31401$ the current does not exist, therefore, flavor pairs have a stable bound, and the embedding  is ME. In general, there exists a $m_{max}$,  where for $m>m_{max}$, the system lives in ME. For $m<m_{max}$, for the small current region, we have both thermal solutions MEH and BE. 
 
As has been already told, this resembles the results in the AdS-Schwarzschild background, except that in the Schr\"odinger solution, we could change chemical potential $\mu$ or $b$.
\noindent As it was previously mentioned, the states in the dual QFT are also labeled with a chemical potential in Schr\"odinger backgrounds, in addition to the temperature. So studying the phase transitions by changing the chemical potential for a fixed temperature will make sense.  
\begin{figure}[h!]
	\makebox[\textwidth][c]{\includegraphics[width=1.0\textwidth]{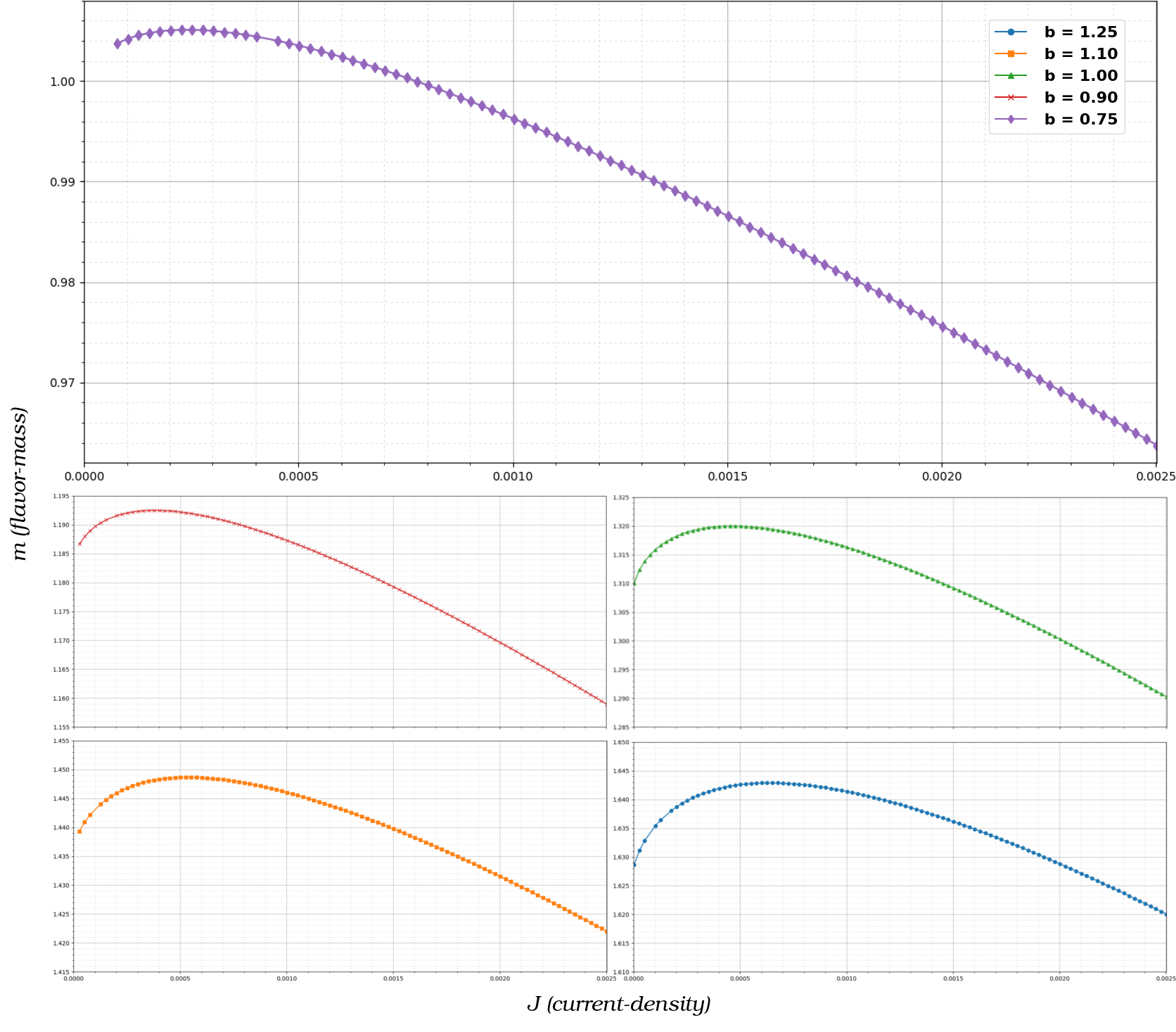}}%
	\caption{$m_q$-$J$ curve for different values of b at $T=0.45015$ and $E_b=0.15$. the maximum values are located at $m_{max}=1.64289$, $1.44865$, $1.32000$, $1.19248$, $1.00509$; from high to low b, respectively.}
	\label{fig:2}
\end{figure}
Hence, both temperature and chemical potential, separately,  could control the stable and non-stable states of different embeddings. These different states in the conducting state, with a different current density and an equal flavor's mass, motivate us to study the phase transition. 

Note that if we recall the relativistic electric field through $E=2\,b \,E_b$, for a fixed value of $E_b$, the lower $b$ results in a smaller $E$, and the maximum mass would take a smaller value relative to the higher $b$ (see Fig.~\eqref{fig:2}). This is quite similar to the AdS result. This means that the chemical potential $\mu$ or $b$ has a major effect on the distances between D3-and D7 branes, or mass of flavors.

\subsection{Negative To Positive Resistivity: The Phase Transition}
\label{sec:PT}
\begin{figure}[h!]
	\makebox[\textwidth][c]{\includegraphics[width=1.2\textwidth]{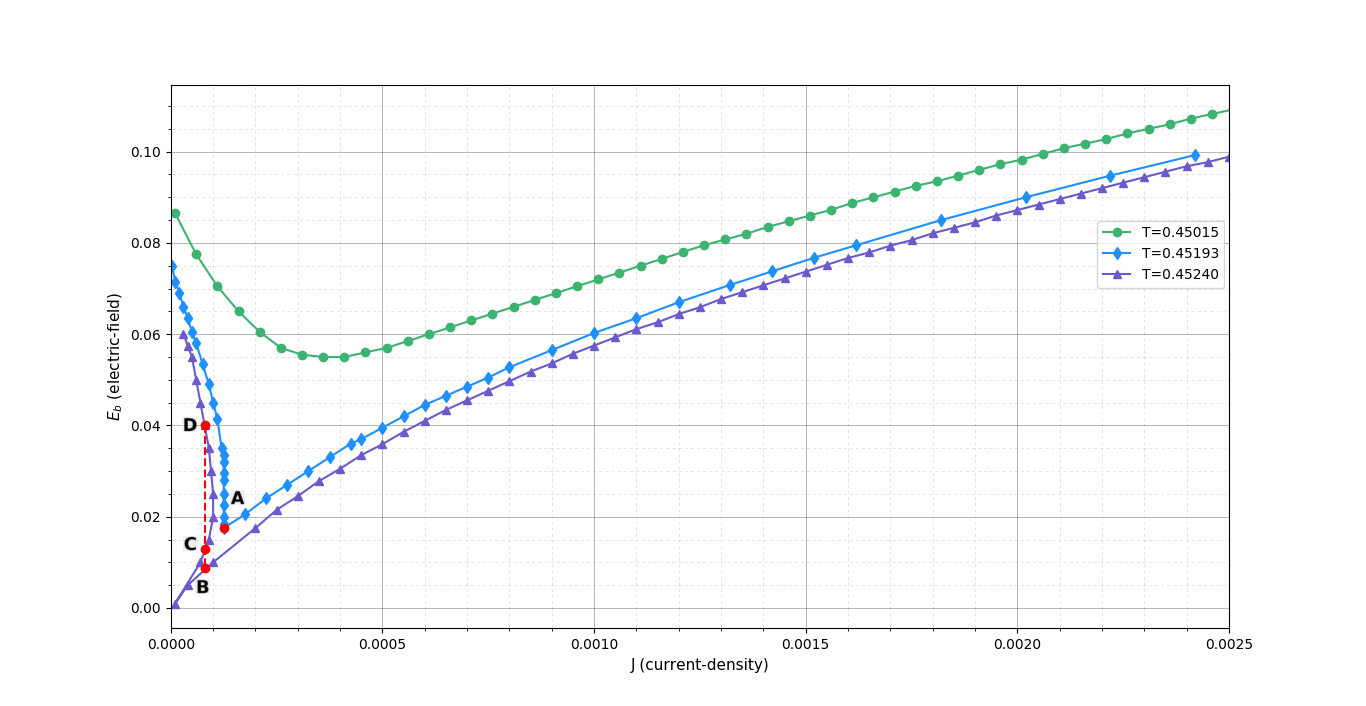}}%
	\caption{$E_b$-$J$ curve for different values of T at $m_q=1.315$ and $b=1$. the $T_c=0.45193$ is the critical temperature at this setting.}
	\label{fig:3}.
\end{figure}

As it was previously pointed, we are dealing with a non-equilibrium steady state, Therefore, in below, we are going to study the non-equilibrium phase transition. By fixing the mass of flavors, we are able to sketch the ($E_b$ -$J$)'s plot.\\
For the different values of temperature and a fixed chemical potential $\mu$ (or $b$), we would have Fig.\eqref{fig:3}.
As it is clear from the Fig.\eqref{fig:3}, For the background temperature $T=0.45015$, the differential conductivity changes continuously by increasing the current density. In the small current region,  we have a negative differential conductivity (NDC), and for the higher currents, we have positive differential conductivity(PDC). This is a continuous change or a crossover, from NDC state to PDC state. Moreover, the electric field has a lower bound, which we call it $E_b^c$. For example, for $T=0.45015$ we have $E_b^c=1.1$. For $E<E_c$ the current is zero which means the Minkowski embedding exists there.\\
For $T=0.45240$, in a region with a small current density, there are points with the same value of currents and different electric fields, i.e., points B,C,D. Since the system lives in a non-equilibrium state, following \cite{Nakamura:10,Nakamura:18}, we define energy or non-equilibrium thermodynamic potential, to find out which one is energetically(or thermodynamically) preferable, as:
\begin{equation}
\mathcal{F}(T,b,J;m_q)=\lim_{\epsilon \to 0}\left(\int_{\epsilon}^{r_H}dr\, \mathcal{H} -\mathcal{H}_c(\epsilon) \right)
\end{equation}
where 
\begin{equation}
\mathcal{H}=-\mathcal{L}_{eff}+\partial _\tau A_x\left(
\frac{\partial \mathcal{L}_{eff}}{\partial\, \partial_\tau A_x}\right) \, 
=\frac{g_{rr}^{1/2}}{r^2} \left(\frac{V}{U}\right)^{1/2}
\end{equation}
and $\mathcal{H}_c(\epsilon)$ is the counter term for renormalization of the Hamiltonian or effective action\footnote{For  holographic renormalization in Schr\"odinger background see~\cite{Taylor1,Guica1}. In here we carefully did the regularization numerically}. In contrary to the AdS \cite{Nakamura:10,Nakamura:12}, and similar to the AdS result in the presence of a constant magnetic field \cite{Ali:13}, the point with largest electric field is favorable. So there would be a jump from the NDC branch (point D)  to the PDC branch. We call this discontinuity of conductivity (or the electric field) first order phase transition.\\
For the temperature $T=0.45193$ we see that there is a point near A that $\frac{\partial E_b}{\partial J}=\,\infty$\footnote{or $\frac{\partial J}{\partial E_b}=0$}. Consequently, the NDC to PDC transition is a continuous, or a second order phase transition. We call this temperature the critical temperature $T_c$. At the critical point, the conductivity has a finite quantity, but the differential conductivity is a singular quantity, which means that between the small current region with a negative differential conductivity (NDC), and the larger current region, with a positive differential conductivity (PDC), a continuous phase transition will occur.

\begin{figure}[h!]
	\makebox[\textwidth][c]{\includegraphics[width=1.2\textwidth]{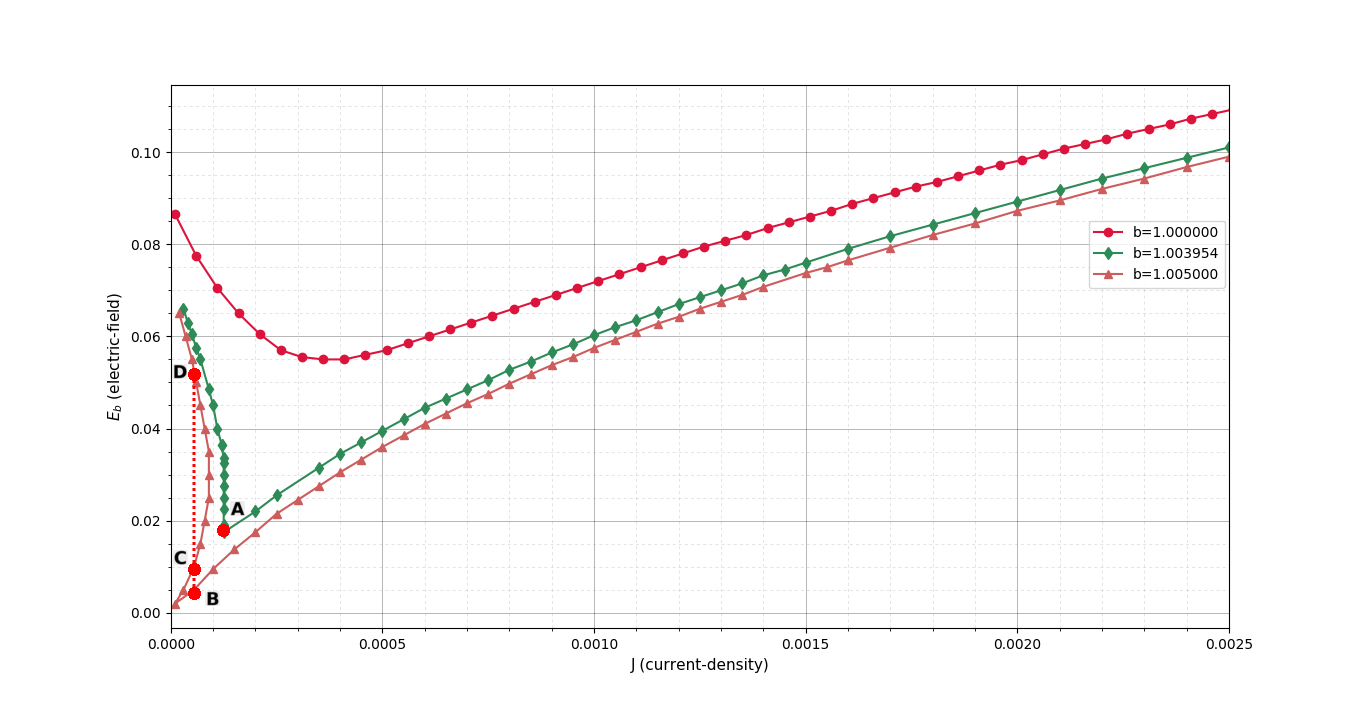}}%
	\caption{$E_b$-$J$ curve for different values of b at $m_q=1.315$ and $T=0.45015$. the $b_c=1.003954$ is the critical b at this setting.}
	\label{fig:4}
\end{figure}

In the case of Schr\"odinger spacetime, we can also change the chemical potential, or b, with a fixed temperature. 
The same behaviour of phase transition occurs due to the variation of the chemical potential, see Fig.\eqref{fig:4}.
For $b=1.003954$ the second order phase transition occurs, where at a point near A, $\frac{\partial E_b}{\partial J}=\,\infty$. Thus we call the chemical potential of that curve:~\textit{critical chemical potential} $\mu_c$ (equally $b_c$). For $b >b_c$, the electric field, hence the conductivity feels discontinuity, so likewise, the first order  phase transition would happen.\\
Concluding from above, we can see that the second order phase transition may occur for different values of $b$ and $T$. We can express this by saying that the critical temperature is controlled by $b$ and vice versa. Therefore, we may have the critical temperature as a function of b , $T_c(b)$, so a \textit{multicritical point} might exist and the system at that point, might belong to a universality class different from normal universality. Due to the numerical difficulties we postpone the study for multicritical point to the future works. In the meanwhile, we can define
\begin{equation}
\tilde{T}=\dfrac{T-T_{ref}}{T_c -T_{ref}}\qquad \& \qquad \tilde{\mu}=\dfrac{\mu-\mu_{ref}}{\mu_c -\mu_{ref}}.
\end{equation}

\begin{figure}[h!]
	\makebox[\textwidth][c]{\includegraphics[width=0.7\textwidth]{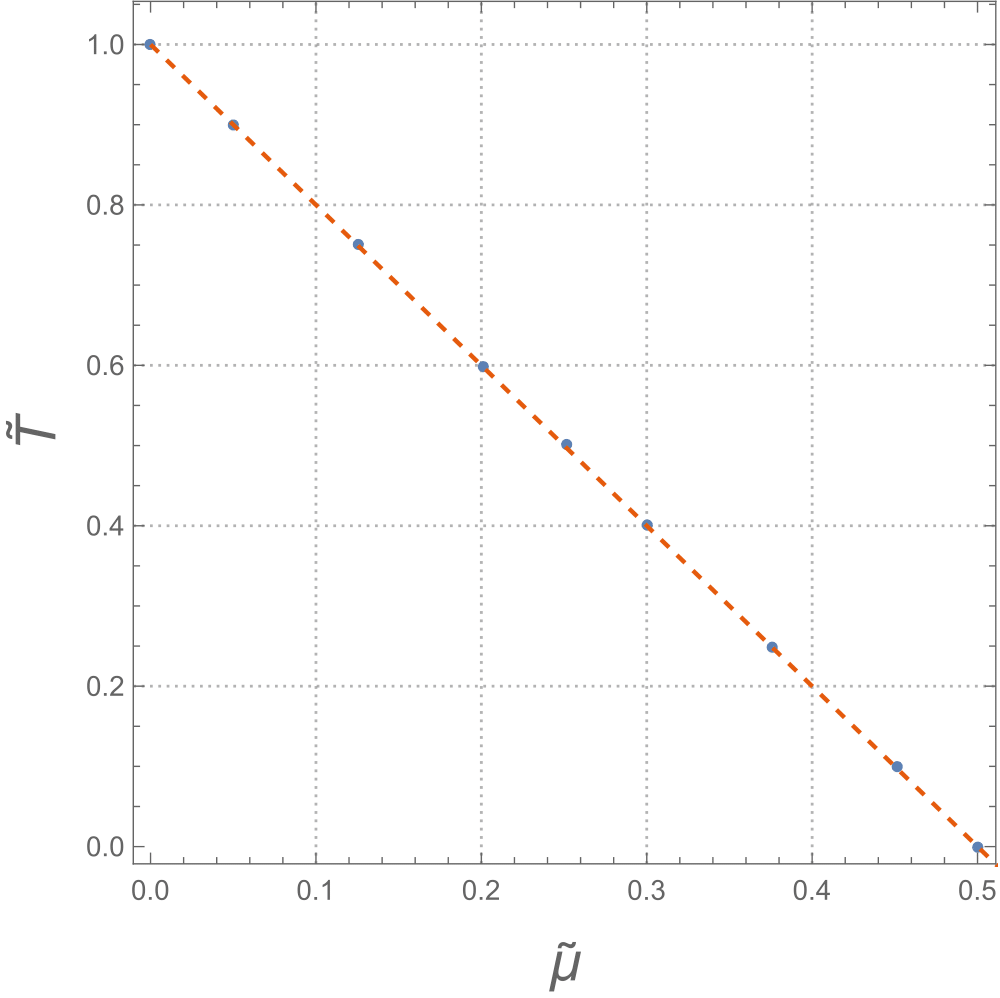}}%
	\caption{Points of different critical states (second order phase transitions) at different $\tilde{T}$ \& $\tilde{\mu}$.}
	\label{fig:7}
\end{figure}
\noindent Using these, we can represent Fig.\eqref{fig:7}, which shows the relation between $\tilde{T}$ and $\tilde{\mu}$. As it was stated above, and from Figure\eqref{fig:7}, by increasing b or chemical potential, the critical temperature decreases. This is quite opposite to the magnetic effect on the critical temperature, see\cite{Ali:13}. 

\subsection{Critical Exponents}
\label{sec:CE}
In the meanwhile of studying the phase transition, we can ask about the scaling behaviour of the order parameters near the critical point. Considering $\sigma-\sigma_c$, along the second order phase transition line $T=T_c$, as an order parameter similar to Landau theory \cite{Nakamura:18}, we can define 
\begin{equation}\label{ce:delta}
\sigma - \sigma_c \propto | J-J_c\,|^{\frac{1}{\delta_{Sch}}} ,
\end{equation}
\begin{figure}[h!]
	\makebox[\textwidth][c]{\includegraphics[width=0.75\textwidth]{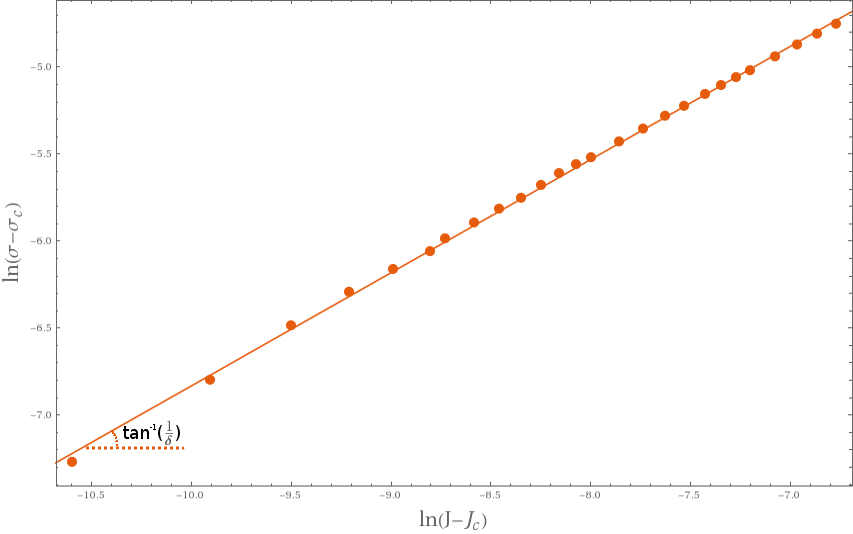}}%
	\caption{behaviour of $ln(\sigma-\sigma_c)$ with the different values of $ln(J-J_c)$, represnting critical exponent of $\delta$ in Schr\"odinger spacetime. $\delta = 1.54 \pm 0.1$ is computed from the slope of the line presented in the figure.}
	\label{fig:5}
\end{figure}
where in here,  $\delta_{Sch} $ is given by the slop of Figure  \eqref{fig:5}. 
\begin{equation}
\delta_{Sch} = 1.54 \pm 0.1
\end{equation}
Interestingly, this is nearly \textit{half} the result in the AdS spacetime~\cite{Nakamura:18}. This observation, makes us propose that dynamical exponent $z$ plays a role in here, and we have in general\footnote{We are starting to study a more general scale invariant theory with dynamical critical exponent $z$~\cite{ADM}.},
\begin{equation}\label{ce:deltaz}
\sigma - \sigma_c \propto | J-J_c\,|^{\frac{1}{\delta _z}}\, ,
\end{equation}
where we have  introduced  
\begin{equation}
\delta _z = \frac{\delta_{AdS} }{z}\, .
\end{equation}
Particularly, this means  that at the critical point, we live in a scale-invariant theory with dynamical critical exponent $z=2$ . Within our numerical resolution, the same behavior of Eq.\eqref{ce:delta} would be predicted along the $b=b_c$ curve, with the same $\delta_{Sch}$.\\
Alongside the first order phase transition lines with $T>T_c$, we can follow \cite{Nakamura:18} to derive the other non-equilibrium critical exponents which has been previously defined as 
\begin{equation}\label{ce:beta}
\Delta \sigma \approx |T -T_c|^\beta \, .
\end{equation}
\begin{figure}[h!]
	\makebox[\textwidth][c]{\includegraphics[width=0.9\textwidth]{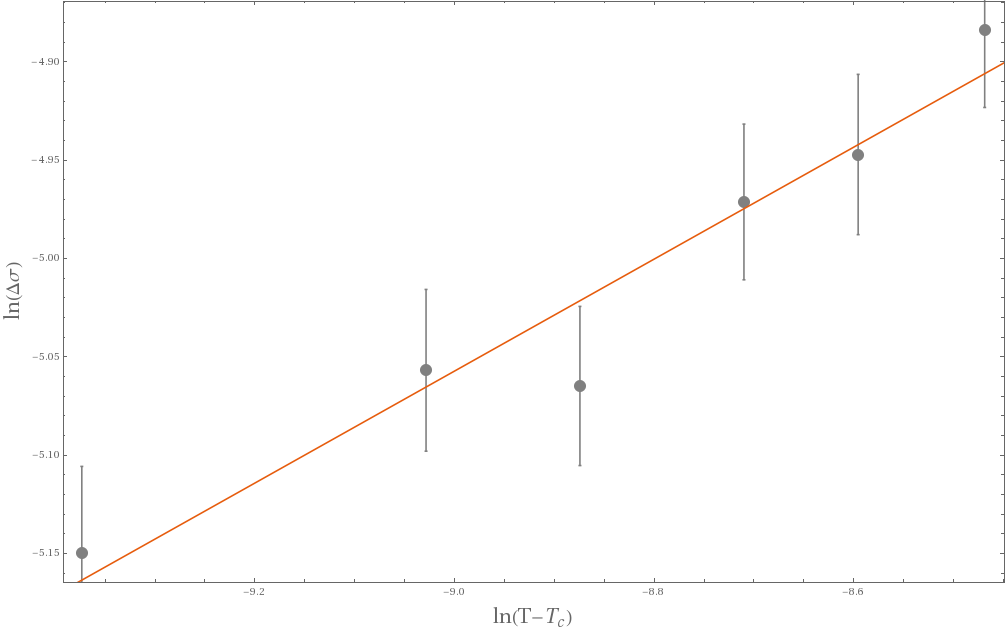}}%
	\caption{behaviour of $ln(\Delta \sigma)$ with the different values of $ln(T-T_c)$, represnting critical exponent of $\beta$ in Schr\"odinger spacetime. $\beta = 0.28 \pm 0.1$ is computed from the slope of the line presented in the figure.}
	\label{fig:6}
\end{figure}
\noindent Interestingly, numerical results fluctuate around half the result which is calculated for AdS one, $\beta=0.28\pm 0.1$. But yet, there exists some level of doubt in the results, due to the numerical difficulties. This result also supports our point that dynamical exponent of $z$ plays an important role in determining the critical exponents. The same argument is true at a constant temperature, for $\Delta \sigma \approx |\mu -\mu_c|^\beta$, from the Fig.\eqref{fig:4}.
\\Results for the critical exponent of $\gamma$, which is also defined in \cite{Nakamura:18}, is postponed to the future works for a further investigation, due to the difficulties arised from the initial definition of this critical exponent.
\\

\section{Probe Branes In Schr\"odinger Spacetime From ALCF: Strange Metal}
Let us start with an $AdS_5\times S^5$, in the unit of AdS radius,
\begin{equation}
ds^2=\frac{1}{r^2}\left(-f(r) dt^2+dx^2+dy^2+dz^2+\frac{dr^2}{f(r)}\right)+d\Omega_5^2
\end{equation}
where $f(r)=1-\frac{r^4}{r_H^4}$ and the $S^5$ metric is
\begin{equation}
d\Omega_{5}^2 = d\theta^2 + \sin^{2}\theta d\psi^2 + \cos^{2}\theta d\Omega_{3}^2.
\end{equation}
The $AdS$ metrics in the light-cone coordinate
\begin{equation}\label{LC}
x^{+}=b(t+y)\qquad and\qquad x^{-}=\frac{1}{2b}(t-y),
\end{equation}
changes to the metrics of ALCF~\cite{Kim:10,Maldacena:2008wh}, which would be 
\begin{align}
ds^2 =\frac{1}{r^2} \left( \frac{1-f }{4b^2 } dx^{+2} - (1+f)dx^+ dx^- + (1-f)b^2 dx^{-2} +  d x^2 +  d z^2 + \frac{1}{f}dr^2\right)+d\Omega_{5}^2 \;.
\label{ALCF}
\end{align}
The event horizon, located at $r_H$, is related to the temperature through $r_H=\frac{1}{\pi \,b \,T}$. This metric interpolates between AdS $z=1$, and $z=2$ Schr\"odinger symmetry~\cite{Kim:10,Kim2:10,Maldacena:2008wh}.\\

\noindent We turn on electric field along $x^+$ as 
\begin{align}
A_+=E_b x ,\,\,\,\,\, A_-=2 b^2 E_b x,\,\,\, A_x=2 E_b b^2 x^- + h_x(r).\label{AAA}
\end{align}
or
\begin{equation}
A_x= -E_b x^+ + h_x(r).
\end{equation}
The Legendre transform of the DBI action is
\begin{equation}\label{leg:ALCF}
S_{D7} = -\int_{0}^{r_H}\!\!dr g_{rr}^{1/2}
\sqrt{U(r)\, V(r)},
\end{equation}
where in here, we have the same $g_{rr}(r)$ as Eq.\eqref{grr}, and
\begin{align}\label{UV:ALCF}
U(r)&=\frac{g_{xx}\big(g_{+-}^2-g_{++}g_{--}\big)-E_b^2 g_{--}}{g_{+-}^2-g_{++}g_{--}}\, \notag\\
V(r)&=\frac{\mathcal{N}^2 f(r)\cos^6\theta}{ r^6}-\lag J^x \rag^2\,\, .
\end{align}
The world-volume horizon located at $r_*$, where both $U$ and $V$ met zero,
\begin{subequations} \label{eq:constrang}
	\begin{equation}
	V (r_*) = 0 \to \lag J^x \rag ^2=\frac{\mathcal{N}^2 f(r_*)\cos^6 \theta(r_*)}{ r_{*}^6}, \label{eq:con5}
	\end{equation}
	\begin{equation}
	U(r_*)=0\to E_b^2=\frac{f(r_*)r_H^4}{\, b^2\,r_*^8}\, .
	\label{eq:con6}
	\end{equation}
\end{subequations}
Therefore, the electric DC conductivity is
\begin{equation}\label{sigma3}
\sigma = \frac{\mathcal{N}\,b\,r_* \cos^3\theta(r_*)}{r_H^2}.
\end{equation}
Due to the electric field dependency of $r_*$ and $\theta(r_*)$, this is a non-linear conductivity. The Eq.\eqref{sigma0} has an extra term, $b^2 f(r_*)\sin^2\theta(r_*)$, comparing to Eq.\eqref{sigma3}.\\
For a non-zero charge density we have to add $h_+(r)$ and $h_-(r)$ to $A_+$ and $A_-$, respectively. In this case we would have
\begin{equation}\label{sigma4}
\sigma =\sqrt{\sigma_1+\sigma_2},
\end{equation}
where $\sigma_1$ is given by the Eq.\eqref{sigma3}, and $\sigma_2\approx\frac{\lag J^+\rag}{g_{xx}(r_*)}$, where $\lag J^+\rag$ is a our charge density~\cite{Kim2:10,kazem1}. It was discussed in \cite{Kim2:10,kazem1}, that when $\sigma_2$ has a dominant effect, the system behaves as a Strange metal. For a zero charge density because $\sigma_2=0$ we might say that the ALCF behaves similar to $z=1$ AdS background.
\\
Same as the Section\eqref{sec:sNMT}, we will study the phase transition numerically, in the conducting state, and investigate its' critical exponents. \\
\subsection{Phase Transition And Critical Exponents}
\begin{figure}[h!]
	\makebox[\textwidth][c]{\includegraphics[width=1.2\textwidth]{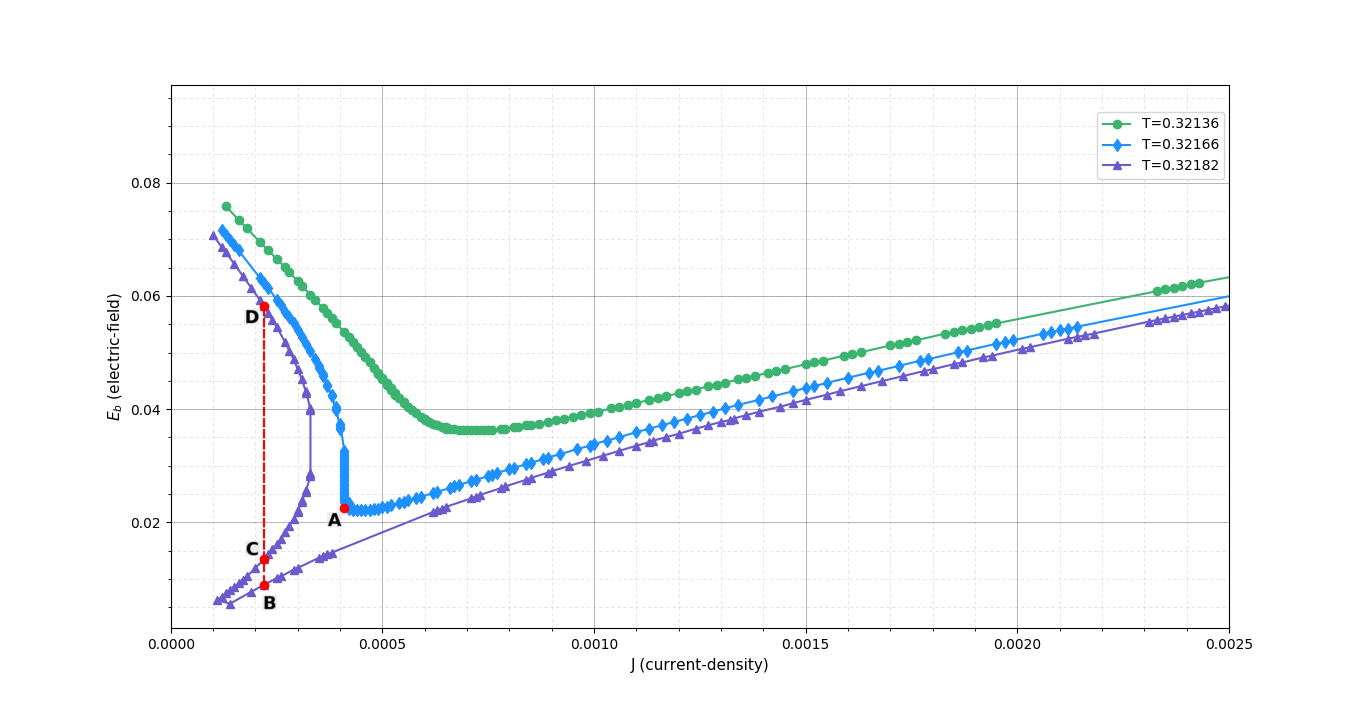}}%
	\caption{$E_b$-$J$ curve for different values of T at $m_q=0.936$ and $b=1$ of the Strange metal. the $T_c=0.32166$ is the critical temperature at this setting.}
	\label{fig:9}
\end{figure}

Following the previous sections' methodology for our Strange metal, our numerical calculations result in the Fig.\eqref{fig:9}. We can see that the same phenomena of phase transition is taking place in here.
It was found that the occurance of phase transition phenomena is not governed by the variation of $b$. So for the ALCF background, the parameter $b$ is showing a quite different characteristics, compared to the previous results in Section\eqref{sec:sNMT}. 
To study the first order phase transition, again, we should compare the values of energy between three points with a same current density in the low current region, i.e., $B,C,D$. The energy density would be
\begin{equation}
\mathcal{H}=-\tilde{\L}+\partial _+ a_x\left(
\frac{\partial \tilde{\L}}{\partial\, \partial_+ a_x}\right) \, 
=\frac{ g_{rr}^{1/2}}{r^2} \left(\frac{V}{U}\right)^{1/2}
\end{equation}
Same as the Schr\"odinger sapcetime from NMT transformation, the point with the higher electric field has a lower energy, hence it is more favorable. This is quite interesting since we'll show that in below, if we turn on the electric field in the light-cone direction, we are getting back the AdS results, which in there, the points with a lower electric field were more favorable.

\begin{figure}[h!]
	\makebox[\textwidth][c]{\includegraphics[width=0.9\textwidth]{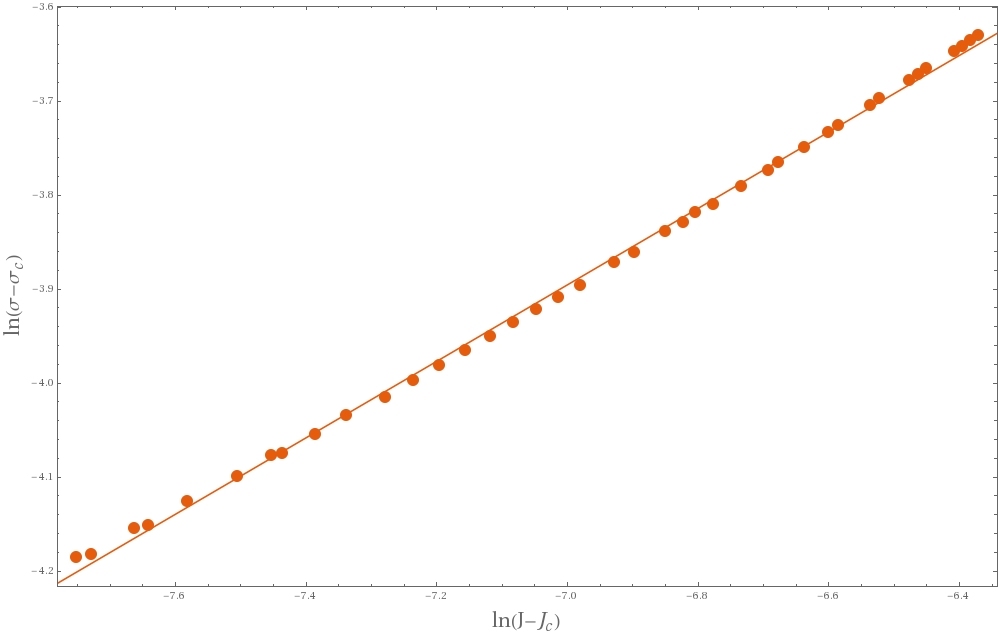}}%
	\caption{Behaviour of $ln(\sigma-\sigma_c)$ with the different values of $ln(J-J_c)$, representing the critical exponent $\delta$ for Strange metal. $\delta = 2.25 \pm 0.2$ is computed from the slope of the line presented in the figure.}
	\label{fig:11}
\end{figure}

\begin{figure}[h!]
	\makebox[\textwidth][c]{\includegraphics[width=1.0\textwidth]{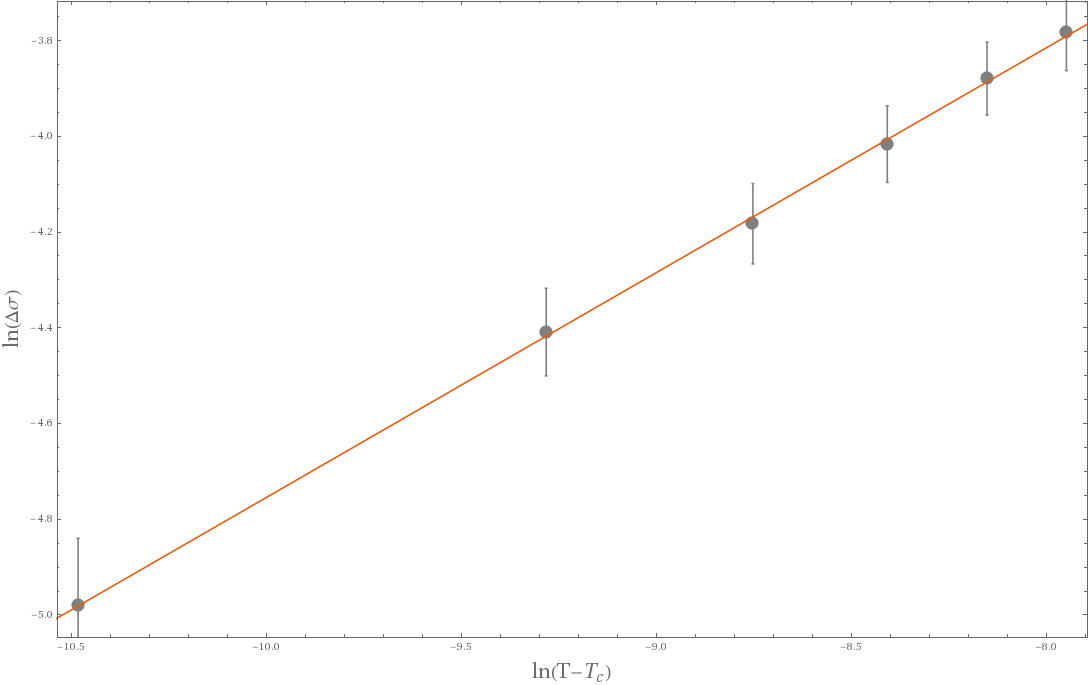}}%
	\caption{Behaviour of $ln(\Delta \sigma)$ with the different values of $ln(T-T_c)$, represnting the critical exponent $\beta$ for strange metal. $\beta = 0.47 \pm 0.05$ is computed from the slope of the line presented in the figure.}
	\label{fig:10}
\end{figure}

\noindent Following \eqref{sec:CE} and in like manner, critical exponents of $\beta$ and $\delta$ are evaluated for our Strange metal. The numerical results of these exponents can be derived from Fig.\eqref{fig:11} and Fig.\eqref{fig:10}. The final result are  $\delta = 2.25 \pm 0.2$ and $\beta = 0.47 \pm 0.05$.
This was expected due to $z=1$ behavior of the system at zero charge density.

\subsection{Electric Field Along The Light-Cone Direction}

The Legendre transformation of the DBI action of the probe $N_f$ $D7$-branes with the gauge field~\eqref{eq:gauge0} would be
\begin{equation}\label{leg:ALCF}
S_{D7} = -\int_{0}^{r_H}\!\!dr g_{rr}^{1/2}
\sqrt{U(r)\, V(r)},
\end{equation}
where
\begin{align}\label{UV:ALCF}
g_{rr}&=\frac{1}{r^2 f(r)}+\theta'^2\notag,\\
U(r)&=\frac{r^4}{f(r)}\left( -4 b^4 g_{++}\, E_b^2 +4 b^2 E_b^2 g_{+-}-E_b^2 g_{--}+\frac{f(r)}{r^6}\right)\ \notag,\\
V(r)&=\frac{\mathcal{N}^2 f(r)\cos^6\theta}{ r^6}-\lag J^x \rag^2\,\, .
\end{align}
The world-volume horizon is located at $r_*$, where both $U$ and $V$ are zero,
\begin{subequations} \label{eq:con}
	\begin{equation}
V (r_*) = 0 \to \lag J^x \rag ^2=\frac{\mathcal{N}^2 f(r_*)\cos^6 \theta(r_*)}{ r_{*}^6}, \label{eq:con3}
	\end{equation}
\begin{equation}
U(r_*)=0\to E_b^2=\frac{f(r_*)}{4\, b^2\,r_*^4}\, .
\label{eq:con4}
\end{equation}
\end{subequations}
Recalling $E=2b\,E_b$ \eqref{eq:con}, these equations will give us the world-volume horizon and also the DC conductivity as same as the AdS results~\cite{KO07}. Hence in here, the phase transition from NDC to PDC is the same as AdS background, see~\cite{Nakamura:12,Nakamura:18}. This is quite trivial since we've made a change of coordinate in the metric and gauge field at the same time, therefore the effective action or DBI action remained unchanged.

 \section{Notes On Numerical Calculation}
\label{sec:NONC}
As it was mentioned earlier, with the increment of the temperature, before reaching the first-order transition, there exists a second-order transition, which contains a single point called critical point, with a diverging $\frac{\partial E}{\partial J}$. Although the accuracy of J-E curves is important in this work, our numerical analysis points out that for the calculation of critical exponents, the precise indication of the critical point plays the most important factor on the precision of our reported critical exponents.\\
Authors believe, due to the high sensitivity of critical exponents to the determination of the positions of critical points, current and previous numerical reports of the critical exponents~\cite{Nakamura:12,Nakamura:18}, are still doubtful, and a better definition for the critical points is required for a further, more precise investigation.
Furthermore, because of the existence of some numerical difficulties, specially in some spacetimes including Schr\"odinger spacetime, a better numerical technique is also demanded to indicate the critical points.

 \section{Conclusion}
In the QFT with $z=2$ Schr\"odinger symmetry, we study the non-equilibrium phase transition by using holographic probe branes. Following~\cite{Nakamura:10,Nakamura:12,Nakamura:18}, using the numerical analysis, we show that the phase transition could occur in the conductor state. For Schr\"odinger solution from type IIB supergravity, we saw that both background temperature and chemical potential (or rapidity) control the phase structure of the transition.\\
We have also seen that the same phenomena happens in the probe branes in Schr\"odinger spacetime from ALCF but interestingly, unlike the solutions from NMT, the phase transition did not take any effect from the variation of chemical potential.\\
Other than this, it was shown that in all of our spacetimes, for the low current density region of the first order phase transitions, the higher electric fields were energetically more favorable. This was particulary interesting for our Strange metal, because when electric field was in the light-cone direction, we were getting back the AdS results, which in there, the points with a lower electric field were more favorable.

\noindent Critical exponents were also calculated for NMT and ALCF. These results, with the addition of the results from \cite{Nakamura:18}, are concluded in the following table.

 \begin{table}[H]
	\centering
	\begin{tabular}{|c|c|c|c|c|c|c|c|}
		\hline
		Critical Exponents&NMT&ALCF&AdS\cite{Nakamura:18} \\
		\hline 
		$\delta$ & $1.54 \pm 0.1$ & $2.25 \pm 0.2$  & $3.008 \pm 0.032$ \\
		\hline
		$\beta$ & $ 0.28 \pm 0.1$ & $0.47 \pm 0.05$ & $0.505 \pm 0.008$ \\
		\hline
	\end{tabular}
	\caption{\label{tab:2}Summary of the calculated critical exponents}
\end{table}

\noindent We propose a relation between the dynamical critical exponent $z$, and $\delta$ and $\beta$ in our Schr\"odinger spacetime from NMT. As it can be seen from Table\eqref{tab:2}, for our $z=2$ NMT, the critical exponents are half the values of AdS ones. Therefore, we propose that $\delta _z = \frac{\delta_{AdS} }{z}$. A further investigation on a more general scale invariant theory with dynamical critical exponent $z$ is carried on~\cite{ADM}.
\\
With the study of the critical exponents, we show that ALCF has a phase structure similar to the relativistic theory from AdS background. From Table\eqref{tab:2}, the difference between the value of $\delta$ for ALCF and AdS is visible. As it is already discussed in Section\eqref{sec:NONC}, this difference might be a result of the numerical inaccuracies in the determination of the critical points. A further investigation is needed for a better definition of the critical points, and regarding the numerical techniques used to precisely determine these points in any spacetime.
 

\end{document}